\documentclass[aps,prd,nofootinbib,twocolumn,superscriptaddress,preprintnumbers,letterpaper]{revtex4-2}
\pdfoutput = 1
\usepackage[T1]{fontenc} 
\usepackage{hyperref}
\usepackage{amssymb}
\usepackage{amsmath}
\usepackage{epsfig}
\usepackage{comment}
\usepackage{color}
\usepackage{multirow}
\usepackage{capt-of}
\usepackage{siunitx}
\usepackage{graphicx}
\usepackage{slashed}
\usepackage{tabularx}
\usepackage{enumitem}
\usepackage{multirow}
\usepackage{booktabs}
\usepackage{isotope} 
\usepackage{xcolor}
\usepackage{bm}
\usepackage{array,mathtools}
\usepackage{diagbox, eqparbox, hhline}

\newcolumntype{C}{>{$}c<{$}}
\AtBeginDocument{
\heavyrulewidth=.08em
\lightrulewidth=.05em
\cmidrulewidth=.03em
\belowrulesep=.65ex
\belowbottomsep=0pt
\aboverulesep=.4ex
\abovetopsep=0pt
\cmidrulesep=\doublerulesep
\cmidrulekern=.5em
\defaultaddspace=.5em
}

\usepackage{tikz}
\usepackage{tkz-euclide}
\usetikzlibrary{backgrounds}
\usetikzlibrary{decorations.pathmorphing}
\usetikzlibrary{arrows.meta}
\tikzset{
    v/.style={decorate, decoration={snake, segment length=2.mm, amplitude=0.5mm}, draw},
    f/.style={draw,decoration={markings,mark=at position #1 with {\arrow[]{Latex[length=1.5mm,width=1.5mm]}}},postaction={decorate},node contents=#1},
    fr/.style 2 args={draw,decoration={markings,mark=at position #1 with {\arrow[rotate=#2]{Latex[length=1.5mm,width=1.5mm]}}},postaction={decorate},node contents=#1},
    f/.default=.6,
    fr/.default={.6}{0},
    fb/.style={draw,decoration={markings,mark=at position #1 with {\arrowreversed[]{Latex[length=1.5mm,width=1.5mm]}}},postaction={decorate},node contents=#1},
    fb/.default=.4,
    fnar/.style={draw},
    g/.style={decorate, draw,  decoration={coil,amplitude=3pt, segment length=3.5pt}},
    s/.style={dashed,draw, postaction={decorate},
        decoration={markings,mark=at position .55 with {\arrow[very thick]{latex}}}},
    sb/.style={dashed,draw, postaction={decorate},
        decoration={markings,mark=at position .55 with {\arrowreversed[draw=black,very thick]{latex}}}},
    snar/.style={dashed,draw,line width =1.25pt},
}

\newcommand{\MM}{\mathcal{M}}

\newcommand{\mvec}[1]{\mathbf{#1}}

\newcommand{\prn}[1]{ \left(  #1 \right) }

\newcommand{\avg}[1]{\left< #1 \right>}

\newcommand{\ord}[1]{\mathcal{O}\left(#1 \right)}

\newcommand{\magn}[1]{\left| #1 \right|}

\newcommand{\al}[1]{\begin{align} #1 \end{align}}
\newcommand{\Msq}{\overline{\magn{\MM}^2}}
\newcommand{\ion}{{\text{ion}}}

\makeatletter

\begin{document}

\title{
Absorption of Sub-MeV Fermionic Dark Matter by Electron Targets
}

\author{Jeff A. Dror}
\email{jdror1@ucsc.edu}
\affiliation{Department of Physics and Santa Cruz Institute for Particle Physics, University of California, Santa Cruz, CA 95064, USA}\affiliation{Theory Group, Lawrence Berkeley National Laboratory, Berkeley, CA 94720, USA}
\affiliation{Berkeley Center for Theoretical Physics, University of California, Berkeley, CA 94720, USA}

\author{Gilly Elor}
\email{gelor@uw.edu}
\affiliation{Department of Physics, University of Washington, Seattle, WA 98195, USA}

\author{Robert McGehee}
\email{rmcgehee@umich.edu}
\affiliation{Leinweber Center for Theoretical Physics, Department of Physics,
University of Michigan, Ann Arbor, MI 48109, USA}
\affiliation{Berkeley Center for Theoretical Physics, University of California, Berkeley, CA 94720, USA}
\affiliation{Theory Group, Lawrence Berkeley National Laboratory, Berkeley, CA 94720, USA}

\author{Tien-Tien Yu}
\email{tientien@uoregon.edu}
\affiliation{Department of Physics and Institute for Fundamental Science, University of Oregon, Eugene, Oregon 97403, USA}

\begin{abstract} 
We study a new class of signals where fermionic dark matter is absorbed by bound electron targets. Fermionic absorption signals in direct detection and neutrino experiments are sensitive to dark matter with sub-MeV mass, probing a region of parameter space in which dark matter is otherwise challenging to detect. We calculate the rate and energy deposition spectrum in xenon-based detectors, making projections for current and future experiments. We present two possible models that display fermionic absorption by electrons and study the detection prospects in light of other constraints. 
\end{abstract}

\maketitle

\section{Introduction}
\label{sec:Intro}
The observation of dark matter (DM) through its gravitational interactions is indisputable evidence of physics beyond the Standard Model. This has motivated experimental efforts to learn about dark matter by searching for its decays, annihilations, self-interactions, and scattering off Standard Model particles. If DM is sufficiently heavy, the scattering off a target material can deposit a detectable amount of kinetic energy in large-volume detectors such as time-projection chambers (TPCs). The energy deposited by the scattering of non-relativistic DM ($\chi$) off a target ($T$) via $\chi T \rightarrow \chi T$ is, at most, ${\cal O} (100)~{\rm keV}$, demonstrating the need for sensitive, low-threshold direct detection experiments.
 
DM direct detection experiments have pushed the limit on the elastic scattering nucleon cross section close to the neutrino floor for weak scale DM masses. However, for masses below a few GeV, DM typically deposits energy below the experimental threshold, significantly impairing experiments' abilities to probe light DM. Thus, the direct detection program has moved towards alternative scattering targets and lower threshold detectors~\cite{Essig:2011nj,Graham:2012su,Essig:2012yx,Essig:2015cda,Hochberg:2015pha,Hochberg:2015fth,Hochberg:2016ntt,Derenzo:2016fse,Hochberg:2016ajh,Schutz:2016tid,Knapen:2016cue,Essig:2017kqs,Budnik:2017sbu,Cavoto:2017otc,Hochberg:2017wce,Knapen:2017ekk,Szydagis:2018wjp,Baryakhtar:2018doz,Griffin:2018bjn,Kurinsky:2019pgb,Hochberg:2019cyy,Trickle:2019ovy,Coskuner:2019odd,Trickle:2019nya,Campbell-Deem:2019hdx,Kozaczuk:2020uzb,Griffin:2020lgd,Trickle:2020oki}. In parallel, novel DM direct detection signals have been proposed which can be constrained by current detectors such as inelastic scattering~\cite{TuckerSmith:2001hy}, bremsstrahlung~\cite{PhysRevLett.118.031803}, exothermic DM~\cite{Essig:2010ye,Graham:2010ca,Bernal:2017mqb}, boosted DM~\cite{Agashe:2014yua,Bringmann:2018cvk}, and self-destructing DM~\cite{Grossman:2017qzw}. These signals are often present in DM models outside the thermal relic paradigm for which there is a range of possible mechanisms that can explain the observed DM relic abundance (see e.g.,~\cite{Griest:1990kh,Carlson:1992fn,Pospelov:2007mp,Hall:2009bx,Cheung:2010gj,Cheung:2010gk,Hochberg:2014dra,Hochberg:2014kqa,Kuflik:2015isi,Pappadopulo:2016pkp,Farina:2016llk,Dror:2016rxc,Dror:2017gjq,DAgnolo:2018wcn,Dery:2019jwf,Kim:2019udq,DAgnolo:2019zkf,Hall:2019rld,Kramer:2020oqi}).

\begin{figure}[t!]
\centering
\includegraphics[width=0.36\textwidth]{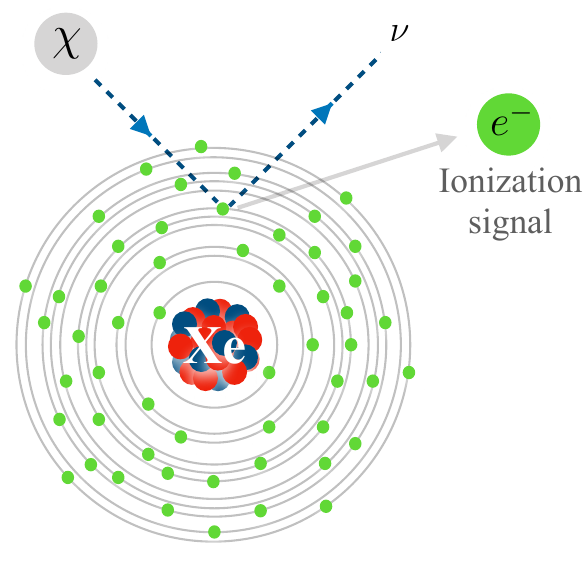}
\caption{Absorption of fermionic DM by electrons in xenon ($A = 54$). A DM particle $\chi$ is absorbed by an electron in a target xenon atom of a direct detection experiment, emitting a neutrino. With enough energy, the electron is ionized and produces a distinctive signal. Fermionic absorption, unlike DM scattering, has a significant contribution from electrons bound to both inner and outer atomic shells.}
\label{fig:Xeabsorb}
\end{figure}

Recently, several of us proposed a class of novel and distinct signatures arising from the absorption of sub-GeV fermionic DM~\cite{Dror:2019onn,Dror:2019dib}. The energy deposited in a fermionic absorption signal is largely independent of the dark matter velocity and parametrically larger than that of DM scattering. Thus, any large-exposure detector can be used to search for this class of signals.~\footnote{The ``inverse'' of fermionic absorption, in which a single DM particle is produced in neutrino scattering, also leads to interesting signals that can be searched for in neutrino experiments~\cite{Chang:2020jwl,Hurtado:2020vlj,Li:2020pfy}.} In \cite{Dror:2019dib}, we considered the specific signals arising from models in which the DM is absorbed by nuclear targets. Such signals can probe DM masses down to an MeV with searches in existing data and significantly below with the proposed lower-threshold experiments. If an atom-bound electron absorbs enough energy from incoming DM to be ionized, the ionized electron may be searched for in photoelectron signatures, known as S2, in TPCs~\cite{Aprile:2019xxb,Akerib:2020lkv,Agnes:2018oej} --- see Fig.~\ref{fig:Xeabsorb} for a schematic of the signal. Current xenon-based direct detection experiments such as XENON1T \cite{Aprile:2019xxb} and LZ \cite{Akerib:2019fml}, as well as future ones such as XENONnT \cite{Aprile:2020vtw}, PandaX-4T \cite{Zhang:2018xdp}, and DARWIN \cite{Aalbers:2016jon}, are sensitive to fermionic absorption of DM with masses in the sub-MeV range by electrons. We explore searches for absorption by electron targets, with a focus on xenon detectors, although our discussion is applicable to other target materials (e.g., liquid argon).

Fermionic absorption by electrons can be induced by vector- and scalar-type operators, given respectively by,
\begin{align}
\label{eq:vectorop}
  O_V &=\frac{1}{\Lambda^2}(\bar\chi\gamma^\mu P_{L,R}\nu)(\bar e\gamma_\mu e)\, ,\\
    O_S &=\frac{1}{\Lambda^2}(\bar\chi P_{L,R}\nu)(\bar e e)\, ,
    \label{eq:scalarop}
\end{align}
where $\Lambda$ may be a ratio of a mediator mass to its coupling. We will consider the case where the mediator is always heavier than the energy transferred during direct detection and hence the interaction is adequately described by an effective operator.

DM capable of inducing fermionic absorption is inevitably unstable. Meta-stable DM changes the equation of state of the universe and can imprint signals in the Cosmic Microwave Background. If the decay products include (or subsequently emit) photons, decaying DM can also be detected with a range of telescopes, strengthening the discovery potential. These searches limit the types of operators as well as the DM mass range detectable by direct detection. In addition, any such operator will have constraints from overproduction near Big Bang Nucleosynthesis (BBN) and searches in colliders which further constrain the viable parameter space. The interplay between the direct detection rates, decay and direct-production bounds is model-dependent. To demonstrate the feasibility of fermionic absorption DM models, we discuss possible UV completions for both operator types. \footnote{For brevity, here and throughout, we use the term ``UV completions,'' although these still require additional states above the weak scale. Up to logarithmic corrections to select multi-loop processes, these states do not contribute to the phenomenology here and we do not specify them further.}

The paper is organized as follows. In Sec.~\ref{sec:eabsorption}, we present the details of the signal arising from fermionic absorption by electrons. Here, we discuss the rate and present the projected constraints at current and upcoming xenon-based experiments. We also briefly discuss and perform a fit to the recent XENON1T excess. The constrained regions of parameter space are model-dependent; so in Sec.~\ref{sec:UVcompl}, we present two UV completions that give rise to fermionic absorption and compute the decay and direct constraint bounds in each case. In addition, we include a model-independent discussion of the irreducible decay constraints. We conclude in Sec.~\ref{sec:Dis}. App.~\ref{app:xsecs} contains a detailed derivation of the differential ionization cross section for fermionic absorption by bound electrons.

\section{Fermionic Absorption by Electrons}
\label{sec:eabsorption}
In this section, we present projected limits at current and upcoming xenon experiments for the two operators in Eqs.~\eqref{eq:vectorop}-\eqref{eq:scalarop} --- the details of possible renormalizable models that generate such operators, along with the associated model-dependent constraints, follow in Sec.~\ref{sec:UVcompl}. We begin by calculating the rate and spectrum of fermionic absorption by (atom-bound) electrons. The results presented here are general and apply to bound electrons of any target material, up to specifying the material-dependent form factors.

Fermionic DM is absorbed by electron targets via the process $\chi+A\to \nu + A^+ + e^-$, where $A$ denotes the atomic number of the target as shown in Fig.~\ref{fig:Xeabsorb}. The underlying interaction is given by 
\begin{equation}
\label{eq:eFabsProc}
\chi+e^-\to \nu+e^-\,,
\end{equation}
which can be mediated by, for instance, the operators of Eqs.~\eqref{eq:vectorop}-\eqref{eq:scalarop}.
In particular, we are interested in the outgoing electron which 
may be detected via scintillation signals or secondary ionizations. Since we are considering the regime where $m_\chi < 2 m_e$ and DM has typical halo velocities $v_\chi\sim 10^{-3}$, the atom is sufficiently heavy so that the target electron is in a static background potential to good approximation. 

\begin{table}[t]
    \centering
    \vspace{0.4 in}
\renewcommand{\arraystretch}{1.5}
\setlength{\tabcolsep}{14pt}
\large{Xenon Binding Energies [eV]}
\begin{tabular}{c@{\hspace{3.5 em}}ccc} \toprule 
\begin{tikzpicture}[every node/.style={inner sep=0,outer sep=0}]
\draw (0,0)--(.33,-.33);
\node at (0.33,0) {$l$};
\node at (0,-.33) {$n$};
\end{tikzpicture}\hspace{-.5cm}&$s$ & $p$& $d$ \\ \midrule
         5 & 25.7 & 12.4 & -- \\
         4 & 213.8 & 163.5 &  75.6\\
         3 & 1093.2 & 958.4 & 710.7\\
         2 & 5152.19 & 4837.7 & -- \\
         1 & 33317.4 & -- & -- \\
         \bottomrule
    \end{tabular}
    \caption{Binding energies $\magn{E_B^{nl}}$ [eV] of the electrons in xenon shells $(n,l)$ calculated from Ref.~\cite{Bunge:1993jsz}.}
    \label{tab:EB}
\end{table}

The momentum of the outgoing neutrino is given by
\begin{equation}
\label{eq:Enu}
p_\nu=\magn{m_\chi \mvec{v_\chi}-\mvec{q}}=\sqrt{m_\chi^2 v_\chi^2+q^2-\!2m_\chi v_\chi q \cos \theta_{qv}} ,
\end{equation}
where $\mvec{v_\chi}$ is the incoming DM velocity, $\mvec{q}$ is the momentum transfer to the target electron, and $\theta_{qv}$ is the angle between $\mvec{v_\chi}$ and $\mvec{q}$, all evaluated in the lab frame. The absorbing electron in the shell $(n,l)$ with binding energy $E_B^{nl}<0$ is ionized with recoil energy $E_R$ (evaluated far from the atom potential). Energy conservation gives
\begin{align}
\label{eq:inelEcons}
& \hspace{-0.2 in} m_\chi+E_B^{nl}-E_R=\\ \nonumber&-\frac{1}{2}m_\chi v_\chi^2+\sqrt{m_\chi^2 v_\chi^2+q^2-2m_\chi v_\chi q \cos \theta_{qv}}\, . 
\end{align}
Typically, the initial momentum of the DM is negligible relative to the momentum of the neutrino, $m_\chi v_\chi \ll q$. This holds when $E_R - E_B^{nl} \ll m_\chi \prn{1-v_\chi}\sim m_\chi$.  Therefore, we can drop $\ord{v_\chi}$ terms and energy conservation simplifies to
\begin{equation}
\label{eq:qfrmEcons}
    q=m_\chi +E_B^{nl} -E_R\, .
\end{equation}
Unless otherwise explicitly stated, all future instances of $q$ will be implicitly dictated by the energy conservation condition in Eq.~\eqref{eq:qfrmEcons}.

Following the procedure discussed in the appendices of \cite{Essig:2015cda} and expanded on in App.~\ref{app:xsecs}, we arrive at the average differential ionization cross section of an electron in the $(n,l)$ shell,
\begin{equation}
\label{eq:sigmaion}
\frac{d\avg{\sigma^{nl}_{\ion} v}}{dE_R} =\frac{\Msq}{64\pi m_\chi m_e^2} \frac{q}{E_R} \magn{f_\ion^{nl} \prn{k',q}}^2 \Theta (q)\,,
\end{equation}
where $\Msq$ is the matrix element squared averaged over initial spins and summed over final ones, and $\Theta$ is the Heaviside step function. $\magn{f^{nl}_\ion \prn{k',q}}^2$ is the ionization form factor of an electron in the $(n,l)$ shell with final momentum $k'=\sqrt{2m_e E_R}$, and is given by~\cite{Essig:2012yx,Essig:2017kqs}
\begin{align}
   &\hspace{-0.1 in} |f_{ion}^{nl}(k',q)|^2=\frac{4k^{\prime3}}{(2\pi)^3}\sum_{l' L}(2l+1)(2l'+1)(2L+1)\nonumber\\
& \qquad \times\left[\begin{matrix} l&l'&L\\0&0&0\end{matrix}\right]^2
 \left|\int r^2dr R_{k'l'}(r)R_{nl}(r)j_L(qr) \right|^2\,,
\end{align}
where $[\cdots]$ is the Wigner 3-$j$ symbol and $j_L$ are the spherical Bessel functions. $R_{nl}(r)$ are the bound electron radial wavefunctions~\cite{Bunge:1993jsz} and the radial wavefunctions $R_{k'l'}(r)$ of the outgoing unbound electrons are obtained by solving the radial Schr{\"o}dinger equation with a central potential $Z_{\rm{eff}}(r)/r$. $Z_{\rm{eff}}$ is determined from the initial electron wavefunction under the assumption that it is a bound state of the same potential. Note that this procedure assumes non-relativistic electron wavefunctions; once $q\gtrsim m_e$, one must take into account relativistic corrections to the electron wavefunctions, which will increase the total cross-section. This enhancement to the cross-section ranges from a factor of a few to several orders of magnitude for $q\sim{\cal O}(\rm MeV)$~\cite{Roberts:2015lga,Roberts:2016xfw}. In our results, we denote the region $m_\chi\gtrsim 100 \text{ keV}$ in which relativistic corrections become important by dashed lines.

The specific form of the differential ionization cross section depends on the underlying operator in the matrix element. For example, for the vector operator in Eq.~\eqref{eq:vectorop} we find
\begin{align}
\label{eq:dsigdERV}
&\hspace{-0.1 in} \frac{d\avg{\sigma^{nl}_{\ion} v}_V}{dE_R} \simeq \frac{\sigma_e }{8 m_e^2 m_\chi^2} \frac{q}{E_R} \magn{f_\ion \prn{k',q}}^2 \Theta (q) \\
&\,\,\times \left[ 2 m_e m_\chi \prn{2 m_e + m_\chi}-2m_e^2 q - q^2 \prn{2 m_e + m_\chi}\right]\, , \nonumber
\end{align}
while the scalar operator in Eq.~\eqref{eq:scalarop} results in
\begin{align}
\label{eq:dsigdERS}
\frac{d\avg{\sigma^{nl}_{\ion} v}_S}{dE_R} &\simeq \frac{\sigma_e q^2}{4 E_R m_\chi^2} \magn{f_\ion \prn{k',q}}^2 \Theta (q)\, ,
\end{align}
where we have kept the terms of ${\cal O}(q^3)$. Note that the operators have a different parametric dependence on $q$. In both cases, $\sigma_e \equiv m_\chi^2/\prn{4 \pi \Lambda^4}$ is a useful parameterization of the cross section.

The differential cross section in Eq.~\eqref{eq:sigmaion} and the corresponding ionization rate are independent of the DM velocity distribution to leading order. As a result, this class of signals is free from the usual astrophysical uncertainties on the local DM velocity distribution that affect direct detection bounds (see {\it e.g. }~\cite{kuhlen2010dark,mccabe2010astrophysical, green2017astrophysical,nunez2019dark,Fairbairn:2012zs,Radick:2020qip}). The total differential ionization rate is found by summing over all possible $(n,l)$ shells of the absorbing target electrons
\begin{equation}
\label{eq:dRiondER}
\frac{dR_\ion}{d E_R} =N_{\text{T}} \frac{\rho_\chi}{m_\chi} \sum_{nl} \frac{d\avg{\sigma^{nl}_\ion v}}{d E_R} \,.
\end{equation}
Here $N_{\text{T}}$ is the number of targets (each of which has all $(n,l)$ shells of bound  electrons) per unit mass and $\rho_\chi \sim 0.4 \text{ GeV}/\text{cm}^3$~\cite{Tanabashi:2018oca} is the local DM energy density. 

Fermionic absorption by electrons can be searched for in a host of experiments and target materials. Specifically, if the incoming DM transfers enough energy to ionize an electron, the resulting ionization signal can be searched for in the S2 data set of TPC experiments \cite{Aprile:2019xxb,Akerib:2020lkv,Agnes:2018oej}. For concreteness, we focus on the absorption of fermionic DM by electrons in liquid xenon, although our discussion is applicable to other noble liquid targets such as argon. 

\begin{figure}[t!]
    \centering
    \includegraphics[width=\columnwidth]{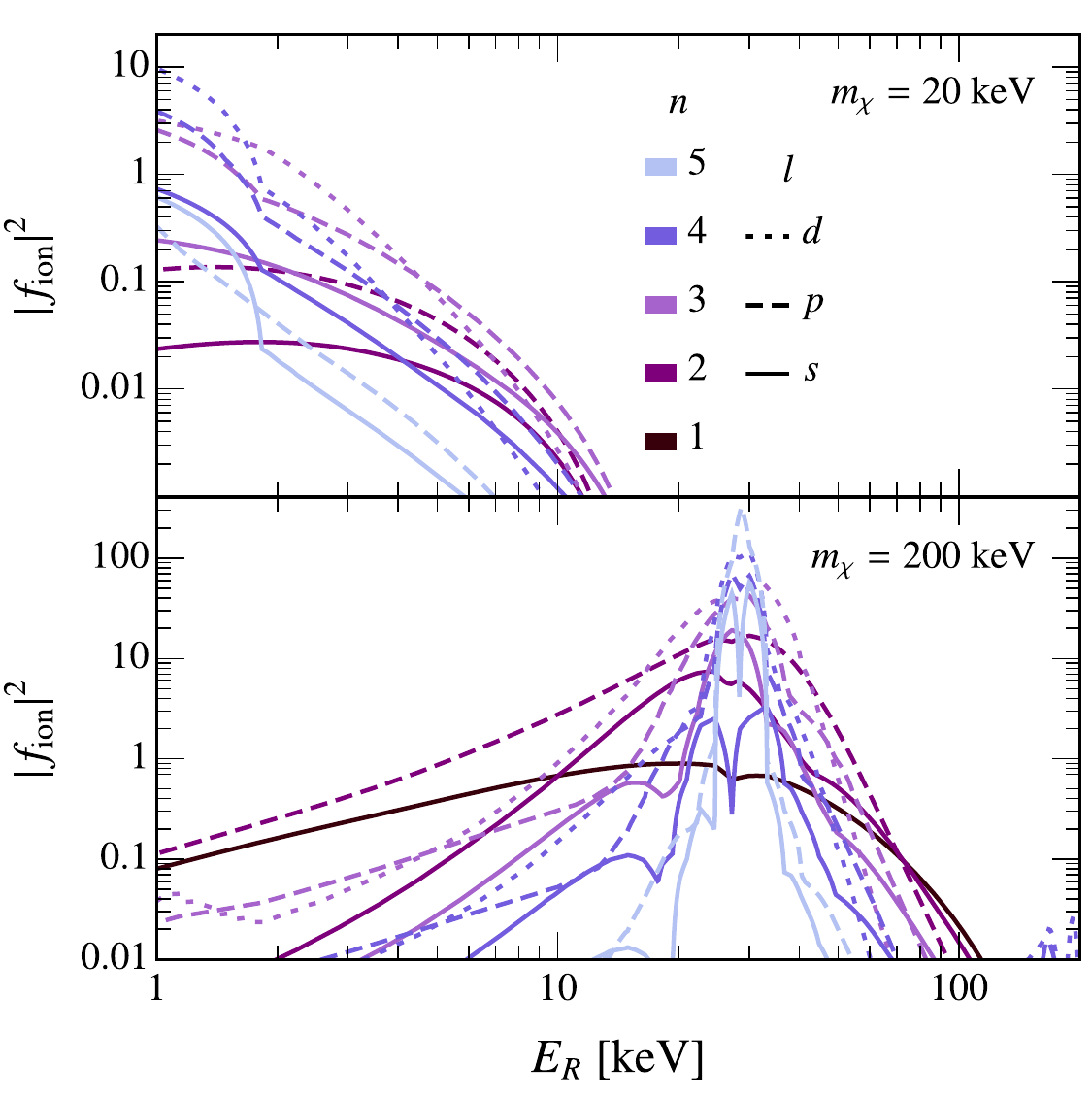}
    \caption{Ionization form factor of each shell in the xenon atom, for $m_\chi=20
    $ keV (top) and 200 keV (bottom). The principle quantum number $n$ is differentiated by color, while the angular quantum number $l$ is differentiated by line-style.
    }
    \label{fig:fion}
\end{figure}

\begin{figure}[t!]
    \vspace{0.3 in}
    \centering
    \includegraphics[width=\columnwidth]{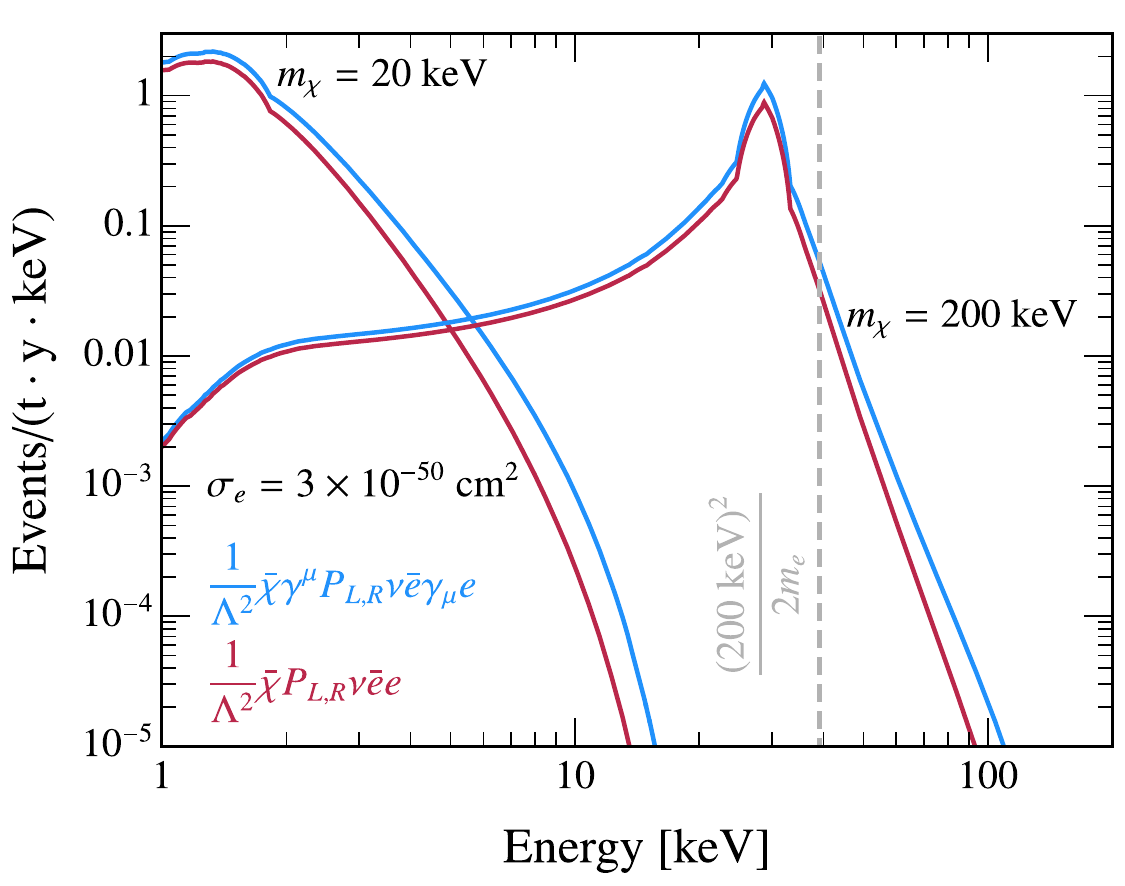}
    \vspace{0.3 in}
    \caption{
    Events rates for $m_\chi=20$ keV and 200 keV at a fixed $\sigma_e=3\times 10^{-50}$ cm$^2$. The rates for the vector (scalar) operator are shown in blue (red). For $m_\chi=200$ keV, we show a dashed gray line representing the expected recoil energy, $E_R=m_\chi^2/2m_e$. For $m_\chi=20$ keV, the DM mass does not solely determine the kinematics of the process and simple intuition breaks down.}
    \label{fig:rates}
\end{figure}

We consider absorption by electrons in the shells with binding energies $\magn{E_B^{nl}}$ shown in Table~\ref{tab:EB}. Note that the kinematics of fermionic absorption necessitate the inclusion of all the electron shells, in contrast to DM-electron scattering which is driven primarily by the outer shell-electrons. In Fig.~\ref{fig:fion}, we plot the ionization form factor of each xenon shell for two benchmark DM masses. For larger DM masses, the incoming electron kinetic energy is small relative to the mass of $\chi$ resulting in form factors becoming localized around $E_R = m_\chi^2/ 2 m_e$, the recoil energy of a free electron absorbing the DM. In Fig.~\ref{fig:rates}, we show a few representative examples of the differential ionization rate in liquid xenon. 
 
\begin{figure*}[t!]
\centering
\includegraphics[width =\columnwidth]{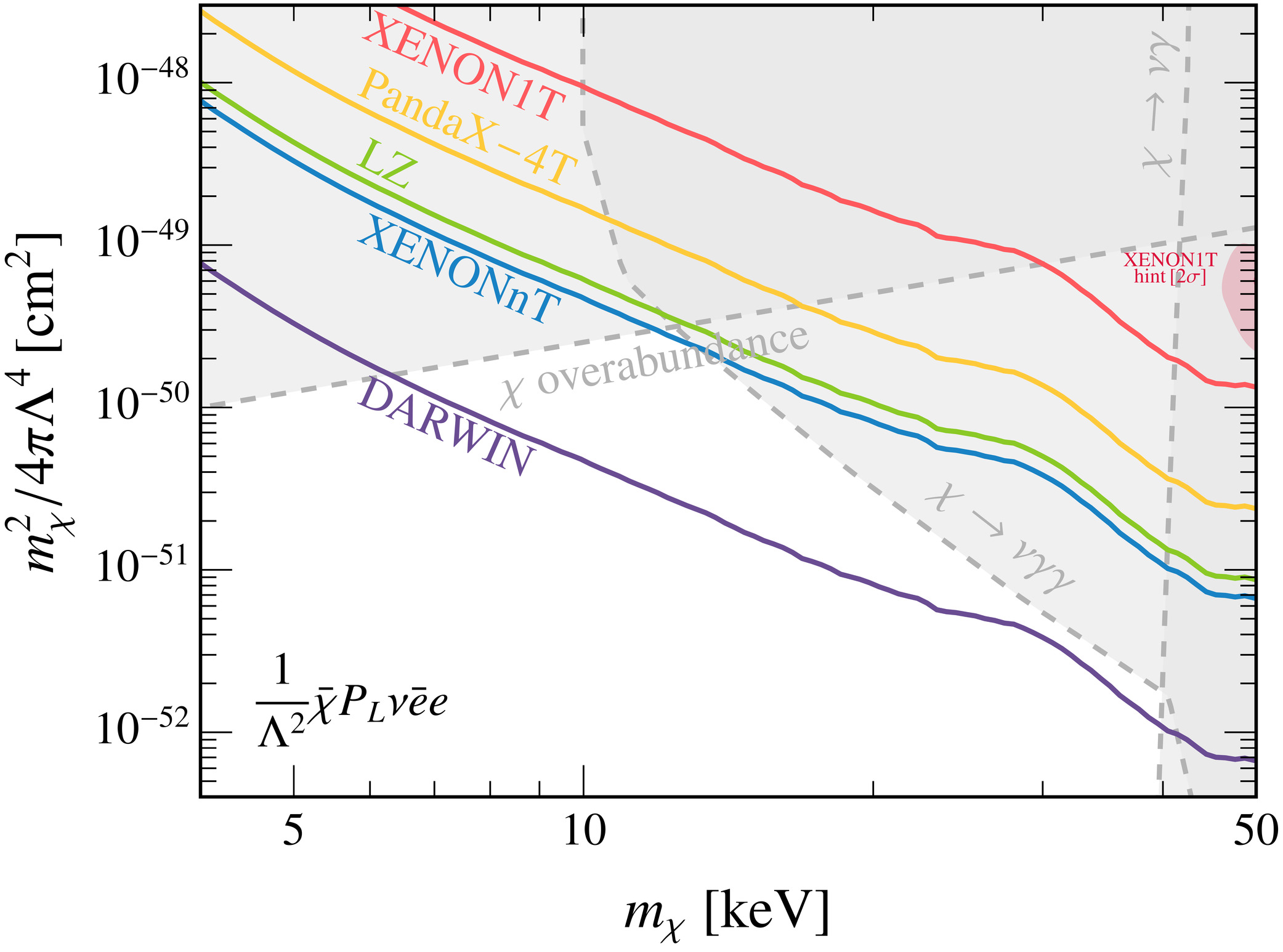}
\includegraphics[width =\columnwidth]{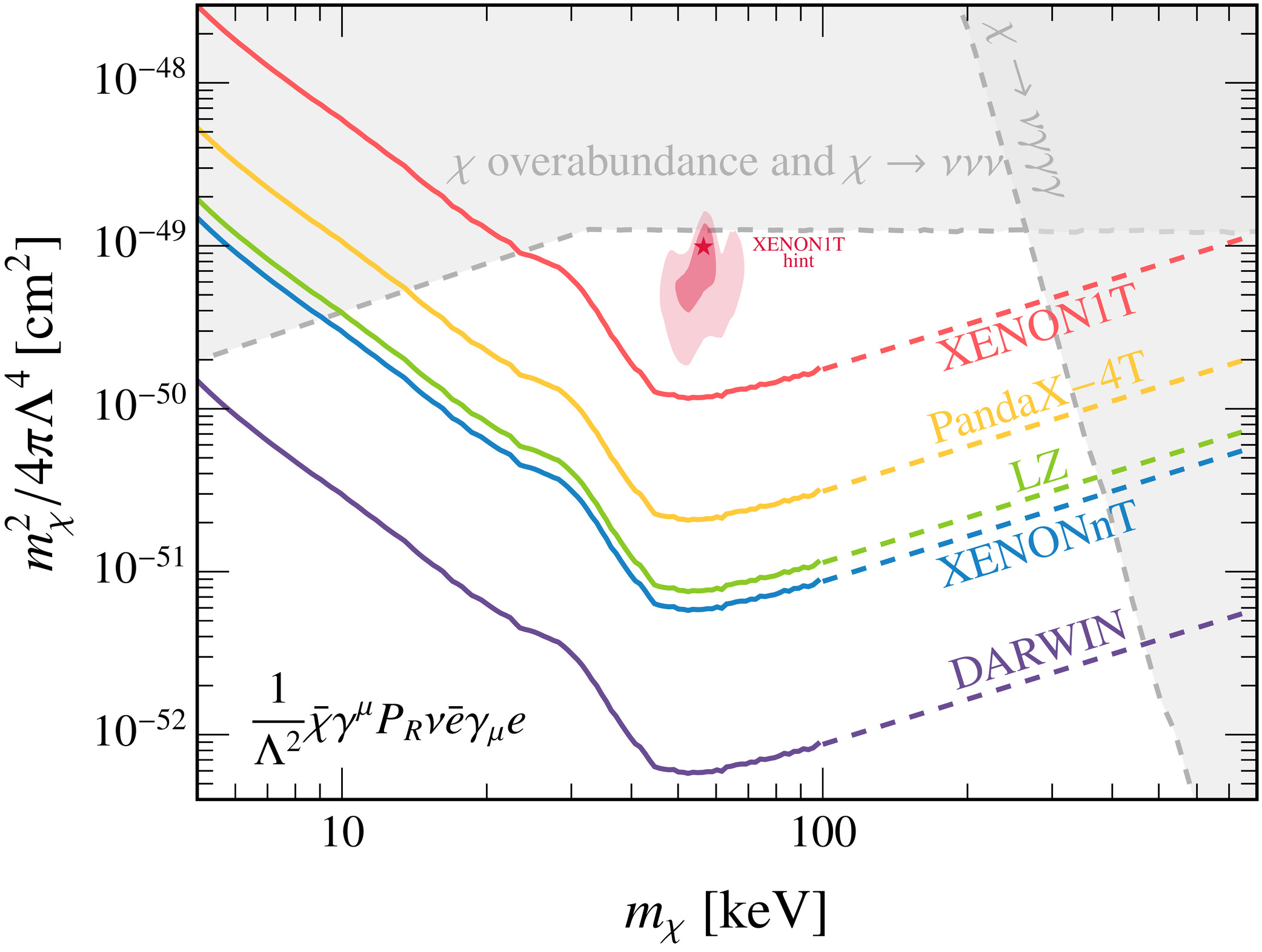}
\caption{
    Projected limits for a signal of fermionic absorption by electrons in current and upcoming xenon target experiments for the scalar ({\bf left}) and vector ({\bf right}) operators. Relativistic effects on the electron wave-function become important in the regions denoted by dashed lines, $m_\chi\gtrsim 100$ keV; these limits are conservative (see text for more details). Other constraints shown in shaded regions are a culmination of bounds from decays, overproduction, and searches for the mediators within specific models. Also shown are the $1\sigma$ and $2\sigma$ best fit regions for the XENON1T excess (red shades) --- the best fit point for the vector model is $(m_\chi, m_\chi^2/4 \pi  \Lambda^4) = (56.5 \, \text{keV}, 10^{-49} \text{cm}^2)$, as indicated by the star.} 
\label{fig:sigmaeproj}
\end{figure*}

We project the sensitivities of XENON1T \cite{Aprile:2019xxb}, LZ \cite{Akerib:2019fml}, PandaX-4T \cite{Zhang:2018xdp}, XENONnT \cite{Aprile:2020vtw}, and DARWIN \cite{Aalbers:2016jon} to the ionization rates in Eqs.~\eqref{eq:dsigdERV} and \eqref{eq:dsigdERS}, corresponding to the operators in Eqs.~\eqref{eq:vectorop} and \eqref{eq:scalarop}, in Fig.~\ref{fig:sigmaeproj}. \footnote{The exposures we assume are: XENON1T (1 t$\cdot$yr), PandaX-4T (5.6 t$\cdot$yr), LZ (15.3 t$\cdot$yr), XENONnT (20 t$\cdot$yr), and DARWIN (200 t$\cdot$yr).} For both operators, the shaded regions are excluded by a combination of indirect detection constraints on DM decay, constraints from the  overproduction of the DM, and direct constraints on the mediator. To calculate the differential ionization rate in XENON1T, we convolve Eq.~\eqref{eq:dRiondER} with their reported total (detector and selection) efficiency using a hard cutoff at $E_R = 1 \text{ keV}$~\cite{Aprile:2020tmw}. We calculate the projected constraints on $\sigma_e$ for various $m_\chi$ by requiring 10 or more events to have occurred over XENON1T's full exposure. We calculate the rates and projections for the other experiments similarly, assuming they have the same efficiency as XENON1T. As discussed above, relativistic corrections to the form factor and absorption rate start becoming relevant at higher $m_\chi$, which we denote by dashed lines. Additionally, the starting assumption that the initial DM kinetic energy and momentum are negligible is only valid when $E_R - E_B^{nl} \ll m_\chi$. Since we impose a hard recoil cutoff at 1 keV, we can only reliably calculate the rates down to $m_\chi \gtrsim 3 \text{ keV}$ and only show this range in Fig.~\ref{fig:sigmaeproj}. \footnote{Detectors with lower electron recoil thresholds may probe DM as light as the phase-space packing bound,~$m_\chi \lesssim 190 \text{ eV}$ \cite{DiPaolo:2017geq,Savchenko:2019qnn}.}

 Existing searches for DM scattering can be recast as limits for fermionic absorption by electrons with the caveat that the kinematics begin to differ for $m_\chi\gtrsim100$ keV due to relativistic corrections of the electron wavefunctions. We find that XENON1T has the potential to probe $20{~\rm keV}\lesssim m_\chi\lesssim 1$ MeV for the dark photon model, but is less effective for the scalar mediator model, as shown in Fig.~\ref{fig:sigmaeproj}. However, LZ, as well as future proposed experiments such as PandaX-4T, XENONnT and DARWIN, have the potential to probe both models. We show bounds from a combination of decays as well as direct and cosmological constraints. For both the vector and scalar cases, the constraints are driven by the over-production of DM and decays. In addition, there are also direct bounds on the mediator which have been taken into account. The decay and direct constraints are sensitive to the models introduced in section~\ref{sec:UVcompl} and may shift depending on the specific UV completion. In contrast, the direct detection signal only relies on the underlying effective operator and is robust given an operator type.

We also explore the possibility that fermionic absorption on electrons can explain the reported XENON1T excess~\cite{Aprile:2020tmw}.~\footnote{It was claimed that a similar model could explain the XENON1T excess \cite{Shakeri:2020wvk}. This work omitted form factors in the matrix elements, which are critical to calculating the absorption rate. In addition, \cite{Shakeri:2020wvk} took DM decay bounds to be of order the age of the universe, while current bounds from indirect detection are significantly more stringent. } 
We preform a $\chi^2$ fit to the XENON1T signal. As a background model, we adopt the ``$\text{B}_0$'' model as presented by the XENON1T collaboration \cite{Aprile:2020tmw}, for which $\chi^2/\rm{dof} = 47.2/29$, and fit the fermionic absorption signal (floating DM mass and $\chi-e$ cross section) plus $\text{B}_0$ background model. The best fit point is $(m_\chi , \sigma_e) = (56.5 \, \rm{keV}, 1 \times 10^{-49} \, \text{cm}^2 )  $  which corresponds to $\chi^2/\rm{dof} = 35.4/27$. The best fit to the XENON1T data is shown Fig.~\ref{fig:sigmaeproj} as a red-star with $1 \sigma$ and $2 \sigma$ contours in mass and cross section. The excess region is easily accommodated by the dark photon UV completion, while for the scalar UV completion, it is firmly excluded by searches for DM decays into $\nu \gamma \gamma$. 

\section{Models}
\label{sec:UVcompl}
We now discuss how the direct detection signal arises in explicit models, effectively resolving the scale $\Lambda$ that gives rise to the higher dimensional operators as in Eqs.~\eqref{eq:vectorop} and \eqref{eq:scalarop}. Importantly, we compute the constraints in each case emphasizing some general features for both the scalar- and vector-mediated models. Constraints will arise from direct searches for the mediators, overproduction of $\chi$ near BBN, and various decays modes. 

\tikzset{mystyle/.style={line width=1,baseline,scale=0.8, every node/.style={scale=0.8}}}
\tikzset{circlestyle/.style={preaction={fill=white},postaction={pattern=north west lines},fill=blue,fill opacity=0.5,draw=black}}
\tikzset{circlestyle2/.style={preaction={fill=white},postaction={pattern=north east lines},fill=red,fill opacity=0.5,draw=black}}
\begin{table*} 
\begin{center} 
\begingroup
\setlength{\tabcolsep}{7.5pt} 
\renewcommand{\arraystretch}{1.25} 
\begin{tabular}{ccccc}
\toprule
{\bf Process:}  & { $ \chi \rightarrow \nu \nu \nu  $}&  $ \chi \rightarrow \nu \gamma $& $ \chi \rightarrow \nu \gamma \gamma  $   & $ \chi \rightarrow \nu \gamma \gamma \gamma $ \\ \midrule
 {\bf 1-loop} &               
 \begin{tikzpicture} [mystyle]
                 \begin{scope}[shift={(0,.5)}]
                   \draw[f] (0,0) node[left]{$ \chi $} -- (1,0);
    \draw[f =0.75] (1,0) -- (2,0.0) node[midway,above] {$ e $};
    \draw[f] (2,.0) -- (3,0)node[right] {$ \nu $};
\draw [fb=0.4 ] (1,0) -- (1,-1) node[below,xshift=0.15cm,yshift=0.1cm] {$ \nu  $};
    \draw[v] (2,0) -- (2,-1) node[midway,right] {$ W $};
    \draw[fb] (2,-1) -- (3,-1) node[right] {$ \nu  $};
    \draw[fb=0.45] (1,0) -- (2,-1) ;
    \draw[circlestyle] (1,0) circle (0.2);
    \node[] at (1,-2) {$+$ crossed};
  \end{scope}
\end{tikzpicture}
 &  \begin{tikzpicture} [mystyle]
                 \begin{scope}[shift={(0,.5)}]
                   \draw[f] (0,0) node[left]{$ \chi $} -- (1,0);
    \draw[f] (1,.0) -- (2,0)node[right] {$ \nu $};
\draw [fr={0.67}{-10}] (1,0) to [out=-155,in=155] (1,-1);
\draw [fr={0.4}{-190}] (1,0) to [out=-25,in=25] (1,-1);
\node at (1.5,-.5){$ e $};
\node at (.5,-.5){$ e $};
\draw[v] (1,-1) -- (1.75,-1.5) node[right]{$ \gamma $};

    \draw[circlestyle] (1,0) circle (0.2);
  \end{scope}
\end{tikzpicture} &\begin{tikzpicture} [mystyle]
                 \begin{scope}[shift={(0,.5)}]
                   \draw[f] (0,0) node[left]{$ \chi $} -- (1,0);
    \draw[f] (1,.0) -- (2,0)node[right] {$ \nu $};
    \draw[f] (1,0) -- (0.5,-0.75);
    \draw[f] (0.5,-0.75) -- (1.5,-.75);
    \draw[f]  (1.5,-.75) -- (1,0);
\draw[v] (.5,-.75) -- (1.25,-1.5) node[right]{$ \gamma $};
\draw[v] (1.5,-.75) -- (1.25+.75,-1.5) node[right]{$ \gamma $};

    \draw[circlestyle] (1,0) circle (0.2);
    \node[] at (1,-2.25) {$+$ crossed};

  \end{scope}
\end{tikzpicture}
& \begin{tikzpicture} [mystyle]
                 \begin{scope}[shift={(0,.5)}]
                   \draw[f] (0,0) node[left]{$ \chi $} -- (1,0);
    \draw[f] (1,.0) -- (2,0)node[right] {$ \nu $};
    \draw[f] (1,0) -- (0.5,-0.75);
    \draw[f] (0.5,-0.75) -- (1.5,-.75);
    \draw[f]  (1.5,-.75) -- (1,0);

\draw[v] (.5,-.75) -- (1.25,-1.5) node[below]{$ \gamma $};
\draw[v] (1,-.75) -- (1.75,-1.5) node[below]{$ \gamma $};
\draw[v] (1.5,-.75) -- (1.5+.75,-1.5) node[below]{$ \gamma $};

    \draw[circlestyle] (1,0) circle (0.2);
    \node[] at (1,-2.25) {$+$ crossed};

  \end{scope}
\end{tikzpicture} \\
 {\bf 2-loop} & 
\begin{tikzpicture} [mystyle]
                 \begin{scope}[shift={(0,.5)}]
                  \draw[f] (0,0) node[left]{$ \chi $} -- (1,0);
    \draw[f =0.75] (1,0) -- (2,0.0) node[midway,above] {$ \nu  $};
    \draw[f] (2,.0) -- (3,0)node[right] {$ \nu $};
\draw [fb=0.4 ] (1,0) -- (1,-1.75)  node[midway,left] {$ e $};
    \draw[v] (2,0) -- (2,-1) node[midway,right] {$ Z $};
    \draw[fb] (2,-1) -- (2,-1.75) node[midway,right] {$ e  $};
    \draw[fb] (2,-1.75) -- (2.75,-1.75) node[right] {$ \nu  $};
    \draw[v] (2,-1.75) -- (1,-1.75) node[midway,above] {$ W $};

    \draw[fb=0.45] (1,0) -- (2,-1) ;
    \draw[circlestyle] (1,0) circle (0.2);
    \draw[f] (1,-1.75) -- (2,-2.25) node[right] {$ \nu $};
    \node[] at (4,-0.75) {$+ \ldots$};

  \end{scope}
\end{tikzpicture}
&  \begin{tikzpicture} [mystyle]
                 \begin{scope}[shift={(0,.5)}]
                   \draw[f] (0,0) node[left]{$ \chi $} -- (1,0);
    \draw[f =.75] (1,0) -- (2,0.0) node[above,midway] {$ \nu  $};
    \draw[f] (2,.0) -- (3,0)node[right] {$ \nu $};
\draw [f =0.75 ] (1,0) -- (1,-1) node[midway,left] {$ e $};
    \draw[f] (1,-1) -- (2,-1);
    \draw[v] (2,0) -- (2,-1) node[midway,right] {$ Z $};
    \draw[v] (1,-1) -- (2,-1.66) node[right] {$ \gamma $};
    \draw[fb] (1,0) -- (2,-1) ;
    \draw[circlestyle] (1,0) circle (0.2);
    \node[] at (4,-0.75) {$+ \ldots$};
  \end{scope}
\end{tikzpicture}& \begin{tikzpicture} [mystyle]
                 \begin{scope}[shift={(0,.5)}]
                   \draw[f] (0,0) node[left]{$ \chi $} -- (1,0);
    \draw[f =.75] (1,0) -- (2,0.0) node[above,midway] {$ \nu  $};
    \draw[f] (2,.0) -- (3,0)node[right] {$ \nu $};
\draw [f =0.75 ] (1,0) -- (1,-1) node[midway,left] {$ e $};
    \draw[f] (1,-1) -- (2,-1);
    \draw[v] (2,0) -- (2,-1.0) node[midway,right] {$ Z $};
    \draw[v] (1,-1) -- (1.75-0.2,-1.66) node[right] {$ \gamma $};
    \draw[v] (1.66,-1) -- (2.33-0.2,-1.66) node[right] {$ \gamma $};
    \draw[fb] (1,0) -- (2,-1) ;
    \draw[circlestyle] (1,0) circle (0.2);
    \node[] at (4,-0.75) {$+ \ldots$};
  \end{scope}
\end{tikzpicture}
 & \begin{tikzpicture}
                 \node at (0,-1){sub-dominant};
                 \end{tikzpicture} \\ \bottomrule
\end{tabular}
\endgroup
\end{center}
\caption[]{Loop-induced diagrams for the leading decays of $ \chi $ induced by a $ \bar{\chi} \Gamma _1 \nu \bar{e} \Gamma _2 e $ operator insertion, denoted by \tikz[]{\draw[circlestyle,line width=1]  circle(2mm);}. Depending on the mediator, $\chi$ may be protected from rapid decays by various symmetries, making certain diagrams vanish without electroweak corrections. We only estimate electroweak corrections parametrically and so do not list all the diagrams. Mixing between $\chi$ and the neutrinos will induce additional decay channels.}
\label{tab:decays}
\end{table*}

Regardless of the details of the UV completion, operators giving rise to fermionic absorption will always lead to some decays computed by inserting these operators into loops of Standard Model states. The list of leading diagrams resulting in DM decay is given in Table~\ref{tab:decays}. Loop-induced decays contain several different scales - an electroweak boson mass, the mass of the mediator making up the direct detection operator, and an electron or $ \chi $ mass. If the mediator mass is above the weak scale, then the loop momentum can often be of order the weak scale, greatly enhancing the induced decay rate. If, on the other hand, the mediator is well below the weak scale (but has small couplings to Standard Model fields to avoid other bounds), then the induced decays are smaller as the loop-momentum can at most be the mediator mass and we will work in this limit. 

DM models with a detectable fermionic absorption rate typically have interactions that induce absorption at leading order and decays at higher coupling and/or loop order. We consider two example UV completions treating the left handed neutrino and electron as independent components. Therefore, above the electroweak scale new states must appear to absorb divergences until renormalization group flow. It is straightforward to extend the models presented here to be fully renormalizable by introducing couplings with the Standard Model Higgs and integrating out additional fields above the weak scale. For the scalar operator, we consider a scalar mediator that couples off-diagonally between DM and a neutrino as a consequence of a global symmetry. For the vector operator, we consider a dark-photon-mediated model with DM-neutrino mixing. In both cases, a robust bound on the scale of the higher-dimensional operator $\Lambda$ comes from requiring that DM is not overproduced in the early Universe. The absorption operator leads to the production of DM between the time of Standard Model neutrino decoupling ($T\simeq 2.3 \text{ MeV}$) and when the electrons leave the bath. This constrains the mediator scale, $\Lambda$, to be above the weak scale for $m_\chi \sim 1 \text{ keV}$, with stronger bounds for heavier masses. In our final results we employ the bounds computed in Ref.~~\cite{Lehmann:2020lcv}.

\subsection{Scalar Mediator}
\label{sec:UVscalarmediator}
In this section, we present a UV completion for the operator
\begin{align} 
& \frac{1}{ \Lambda ^2 } \bar{\chi} P _L  \nu \bar{e}  e +{\rm h.c.} \,.
\label{eq:effop2}
\end{align} 
Consider a theory with a Dirac DM candidate, $\chi$, and a scalar field, $\varphi$. We impose a global $U(1)$ symmetry with charges, $Q_i$, such that $Q_{\chi_L}=-Q_{\chi_R}=+Q_\nu$ with $\varphi$ and the rest of the Standard Model remaining uncharged. Here we assume neutrinos are Majorana fermions and neglect corrections proportional to their mass. The interaction Lagrangian is
\begin{equation} 
{\cal L} \,\,\, \supset \,\,\, y _{ e e } \varphi \bar{e} e  + y _{ \chi  \nu } \varphi \bar{\chi} P _L  \nu +{\rm h.c.} \,.
\end{equation} 
We note that the symmetry is explicitly broken by electroweak interactions such that terms like $\varphi \bar{\nu}P_L \nu$ will be generated at loop order; we estimate the size of such contributions below. Integrating out the scalar results in the operator that gives rise to the fermionic absorption signal in Eq.~\eqref{eq:effop2}, where we identify the higher dimensional operator scale as,
\begin{align}
\label{eq:DDopScalar}
& \frac{1}{ \Lambda ^2 } \equiv \frac{ y _{ \chi  \nu }  y _{ee }  }{ m _\varphi ^2 } \,.
\end{align} 

The dominant $ \chi $ decay modes are shown in Table~\ref{tab:decays} with the operator insertion resolved by an internal $\varphi$ line. Without electroweak corrections, $\chi \rightarrow \nu \gamma $ vanishes and so the leading decay is $\chi \rightarrow \nu \gamma\gamma$, the rate for which is given by
\begin{align} 
\label{eq:scalarnugg}
\Gamma _{ \nu \gamma \gamma  } 
& = 10 ^{-26} {\rm sec}^{-1} \left(\frac{m_\chi}{20~{\rm keV}} \right)^{7} \left(\frac{{\rm TeV} }{\Lambda} \right)^4 \,. 
\end{align} 
Importantly, the rate depends only on $m_\chi$ and $\Lambda$ and not on any other free parameters of the model. This makes the decay a largely irreducible constraint independent of the details of the model and a generic prediction for a scalar-mediated absorption signal. The bound on the decay rate, which is $ {\cal O} ( 10 ^{ - 27} {\rm sec} ^{-1} ) $, is computed over the mass range of interest in Ref.\cite{Essig:2013goa}. 

Other decays arise at higher loop order and are more sensitive to variations in the model parameters. The decay rate for $ \nu \gamma $ is induced by weak interactions and the precise form of the rate depends on the UV completion above the EW scale. Due to this inherent uncertainty, and difficulties in carrying out the two-loop calculation, we settle for an order of magnitude estimate for this process. In the limit that $m_\varphi \ll m_W$, the rate is of order,
\begin{align}
    \Gamma_{\nu \gamma} \sim  10^{-26} {\rm sec}^{-1} \left(\frac{m_\chi}{30~{\rm keV}}\right) \left(\frac{m_\varphi}{{\rm GeV}}\right)^4 \left(\frac{\rm TeV}{\Lambda}\right)^4 \,.
\end{align}
Although this interaction is generated at two-loop order, it is still significant as it is not suppressed by high powers of $m_\chi$. For the limits we use the recast the bound computed in Ref.\cite{Essig:2013goa} (also of $ {\cal O} ( 10 ^{ - 27} {\rm sec} ^{-1} ) $) for a scalar dark matter decaying into two photons.

Decays of $\chi$ into $ 3 $ neutrinos are mediated by weak insertions. In the limit of massless neutrinos, the Lorentz structure of the scalar operator leads to the decay vanishing at one loop but it is generating at two loops leading to an effective dimension-6 operator between $\chi $ and 3 neutrinos, $\bar{\chi}P_L \nu \bar{\nu }P_L \nu$ with a scale,
\begin{equation}
\frac{1}{ \Lambda _{ 3 \nu } } \sim \frac{ y _{ ee } y _{ \chi \nu } m _e }{  (4\pi)^4 m _W ^2 } \frac{ g ^4  }{ 4 }\, , 
\end{equation} 
leading to a decay rate,
\begin{equation}
\Gamma _{ 3 \nu } \sim \frac{ m _\chi  ^3 }{ \Lambda _{ 3 \nu } ^2 } \frac{1}{ (4\pi)^3 }\, .
\end{equation}
Since the invisible decay is poorly constrained relative to visible bounds (see, e.g. Ref.\cite{DES:2020mpv}), it does not end up being significant in any region of the parameter space of interest. 

The bounds on this model from DM decay are shown in the left panel of Fig.~\ref{fig:sigmaeproj}. We see that the robust bounds of the decay $\chi\to\nu\gamma\gamma $ limit the viability of observing fermionic absorption for dark matter heavier than around 50 keV.

\subsection{Dark Photon Mediator}
\label{subsec:DarkPhotonMediator}
We now present a UV completion for the vector operator, 
\begin{align} 
& \frac{1}{ \Lambda ^2 } \bar{\chi} \gamma _\mu P_R \nu \bar{e} \gamma ^\mu e \,. 
\label{eq:effop}
\end{align} 
Consider a Dirac fermion, $\chi$, charged under a dark gauge group, $U(1) '$, with a kinetic mixing, $\varepsilon$, with the Standard Model photon. For energy transfers well below the dark photon mass, $m_{A'}$, there is an interaction between the dark current, $J'_\mu \equiv g_X \bar{\chi}\gamma_\mu \chi$, and the electromagnetic current, $J_\mu$,
\begin{equation} 
  {\cal L}_{\rm eff} \,\,\, \supset  \,\,\,   \frac{\varepsilon}{ m _{A'} ^2 }J '_\mu  J ^{ \mu} \,.
\end{equation} 
At this point, $ \chi $ is stable as a consequence of its $U(1)' $ charge and one cannot generate the effective operator~\eqref{eq:effop}. We now introduce a $U(1) ' $-charged scalar, $ \varphi $, that breaks the symmetry but is neutral under $SU(3)_C \times U(1)_Y $. Furthermore, we focus on the least constrained case where neutrinos are Dirac and $\varphi$ couples to the right-handed component.\footnote{If the fermionic absorption operator is induced by mixing of $\chi$ with the left-handed neutrinos, then there are stringent bounds from $\chi \rightarrow \nu \gamma$.} The $\chi$ mass and interaction Lagrangian is,
\begin{equation} 
{\cal L}  \,\,\, \supset  \,\,\, m _\chi \bar{\chi} \chi + y  \varphi \bar{\chi} P _R \nu  +{\rm h.c.} \,.
\end{equation} 
After $\varphi$ gets a vacuum expectation value  $\left\langle \varphi\right\rangle$, the neutrino mass is largely unchanged\footnote{An easy way to see this is to note that, in the limit that the neutrino mass vanishes, the right-handed neutrino does not have a Dirac-partner, but lepton number is still conserved.}, however the eigenstates shift as
\begin{equation} 
\chi _R \rightarrow \chi _R- \frac{ y \left\langle \varphi \right\rangle }{ m _\chi } \nu _R\,, \qquad \nu _R \rightarrow \nu _R - \frac{ y \left\langle \varphi \right\rangle }{ m _\chi }  \chi _R \,.
\end{equation} 
The $ \chi $ mass remains approximately unchanged if $ y \left\langle \varphi \right\rangle \ll m _\chi $, but the shift induces a new term in the potential. Defining $ \theta \equiv y \left\langle \varphi \right\rangle  / m _\chi $, the shifted Lagrangian in the limit where $\theta$ is small is,
\begin{align}
\nonumber
{\cal L}  \,\,\,& \supset  \,\,\,   \frac{ \varepsilon }{ m _{A'} ^2 } J'_\mu J^\mu    -  \frac{  \varepsilon g_X   \theta  }{ m _{A'} ^2 } J   ^\mu \, \bar{\chi} \gamma _\mu P  _R \nu  \\ 
& -  \frac{2 g _X ^2 \theta ^3}{ m _{A'} ^2 }  \left( \bar{\chi} \gamma _\mu P _R \nu \right) \left(  \bar{\nu} \gamma ^\mu P _R \nu \right)  +{\rm h.c.} +... \,.
\label{eq:Laftershift}
\end{align}  
where the ellipses denote higher order $\theta$ corrections that have no bearing on the phenomenology. The fermionic absorption operator of Eq.~\eqref{eq:effop} (in addition to couplings to the rest of the fermions making up the electromagnetic current) is generated via the identification  
\begin{align} 
& \frac{1}{ \Lambda ^2 } \equiv -  \frac{  g _X e \varepsilon \theta  }{ m _{A'} ^2 } \,. \label{eq:lambda}
\end{align} 
Note that the third term generated in Eq.~\eqref{eq:Laftershift} couples the DM to neutrinos; the result of which is to admit additional DM decay channels which we take into consideration in what follows. 

\begin{figure}[t!]
    \centering
    \includegraphics[width=\columnwidth]{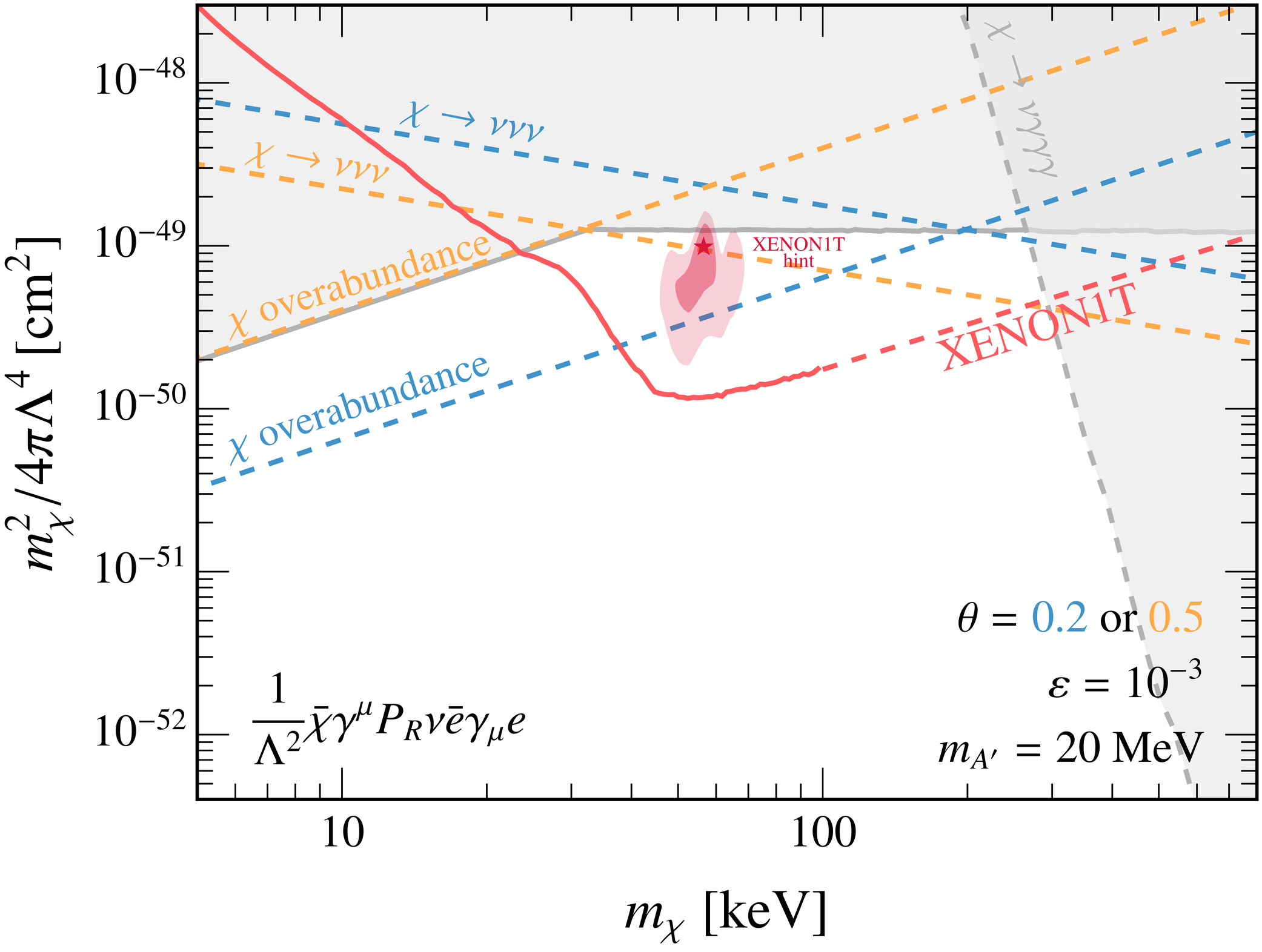}
    \caption{We show the direct and decay constraints in the dark photon model for two benchmark values of the DM-neutrino mixing angle, $\theta = 0.2$ (blue) and $ \theta = 0.5 $ (orange). The gray region corresponds to the irreducible constraints on the dark photon model as described in the main text.}
    \label{fig:DarkPhotonBenchmark}
\end{figure}

The leading decay channels for the vector-mediated model are shown in Table~\ref{tab:decays}. Since $\chi$ mixes with the right-handed neutrino, the decay of $\chi \rightarrow \gamma \nu $ requires a neutrino mass insertion and can be negligibly small. The decay of $\chi$ into $\nu\gamma\gamma$ does not arise at 1-loop since loops of vector-like fermions with an odd number of external vector legs are zero by charge conjugation symmetry. The leading decays are thus $\chi\rightarrow 3 \nu$ and $\chi \rightarrow \nu \gamma \gamma \gamma $. The decay rates for these were computed for a similar model with dark photons coupled to nucleons in \cite{Dror:2019dib} and we repurpose them here. Given in terms of model parameters, the decay rates are  
\begin{align} 
\Gamma _{ \chi \rightarrow \nu \gamma \gamma \gamma } &\,\, \simeq \,\, 10 ^{ - 26}{\rm sec} ^{-1} \left( \frac{ m _\chi }{ 320~{\rm keV} } \right) ^{13} \left( \frac{ {\rm TeV} }{  \Lambda } \right) ^4\,,  \\
\Gamma _{ 3 \nu }   &\,\, \simeq \,\, 10 ^{ - 17} {\rm sec} ^{-1}  \left( \frac{ m _\chi }{ 20~{\rm keV} } \right) ^5 \left( \frac{ m _{A'} }{ {\rm GeV} } \right) ^4 \\ \nonumber
&\qquad \quad \,\, \times \, \left( \frac{ 10 ^{ - 3} }{ \varepsilon } \right) ^4  \left( \frac{ {\rm TeV} }{ \Lambda } \right) ^8 \left( \frac{ \theta }{  0.01} \right) ^2 \,. 
\end{align} 
For setting limits on the $ \chi \rightarrow \nu \gamma \gamma \gamma  $ decay, we approximate the bound with those computed in Ref.\cite{Essig:2013goa} for dark matter decaying to an invisible state and two photons. For setting limits on $ \chi \rightarrow 3 \nu $ decay channel, we use the results from Ref.~\cite{DES:2020mpv} which restrict $ \Gamma _{ 3  \nu } ^{-1} \gtrsim 50~{\rm Gyr} $. The $\nu \gamma \gamma \gamma$ decay will be present with (at least) this rate for any model that leads to a vector-mediated operator and can be considered as an irreducible decay; it may be possible to construct models which evade the $3\nu$ channel. 

In addition to decays, direct dark photon searches place constraints on the parameter space. Constraints on dark photons are well documented and have been summarized recently for visible decays in {\it e.g.} Fig. 4 of Ref.~\cite{Ilten:2018crw}. In making plots, we fix the dark photon mass and mixing angle to be $m_{A'} = 20~{\rm MeV}$ and $\varepsilon=10^{-3}$, leaving the remaining free parameters as $g_X$, $m_X$, and $\theta$. The resulting projected sensitivity of searches for the dark photon mediator model are presented in Fig.~\ref{fig:sigmaeproj}. The decay and direct constraints depend on the value of the mixing angle. In Fig.~\ref{fig:DarkPhotonBenchmark}, we show the limits for two benchmark values of $\theta$. The shaded bands in Fig.~\ref{fig:sigmaeproj} correspond to the minimal decay and direct constraints (the intersection of decay and direct bounds as $\theta$ varies) such that there exists a value of $\theta$ for which the point in the parameter space is otherwise allowed. 

\section{Discussion}
\label{sec:Dis}
In this work, we introduce a new class of signals where fermionic DM is absorbed by electrons and present models which give rise to such signals; these are a natural extension of Fermionic DM Absorption~\cite{Dror:2019onn,Dror:2019dib}. For concreteness, we have focused primarily on targets in liquid xenon and XENON1T's capabilities to discover these signals. In addition, we calculated the projected constraints for PandaX-4T, XENONnT, LZ, and DARWIN, assuming similar detector efficiencies. We have found XENON1T, with its current exposure, can probe a DM of mass $20{~\rm keV} \lesssim m_\chi\lesssim 1$ MeV for the dark-photon-mediated model presented here, while future experiments can go down to a few keV in DM mass. Although XENON1T is not sensitive to the scalar-mediated model, near-future experiments can probe the scalar model for DM masses of about $10{~\rm keV}\lesssim m_\chi\lesssim 40{~\rm keV}$. Next-generation detectors, such as DARWIN, will be able to probe orders of magnitude more parameter space in both cases. Argon-based detectors, such as DarkSide-50~\cite{Agnes:2018ves}, DarkSide-20K~\cite{Aalseth:2020nwt}, and Argo~\cite{Aalseth:2017fik}, will give rise to similar constraints for comparable exposures and detector thresholds; we leave these calculations to future work. 

While we focused entirely on direct detection experiments, the potentially large energy deposits of fermionic absorption open the possibility of detection in neutrino detectors, if they have energy thresholds below an MeV. A notable neutrino experiment with a sufficiently low energy threshold is the CUORE experiment~\cite{Alduino:2017ehq}. Although built to search for neutrinoless double-beta decays,  it can achieve a threshold of a few keV~\cite{DiDomizio:2010ph} with exposures comparable to XENON1T. For heavier $\chi$, experiments such as Borexino~\cite{Agostini:2018fnx} can leverage their huge exposure to search for fermionic absorption. While Borexino has an electron recoil energy threshold of $\sim 70~\text{keV}$, its exposure is $\sim 817$ times that of XENON1T, making it much more sensitive to absorption for heavier DM $m_\chi \gtrsim 100~\text{keV}$. Computing the ionization form factors of different targets would allow one to calculate the projected absorption rates in various other experiments, which is beyond the scope of this work. 

Relaxing the assumption of DM stability leads to a new set of DM direct detection signals that can be probed by XENON1T and future liquid xenon experiments. This offers a new opportunity to probe DM with masses below an MeV, a region generally experimentally inaccessible for elastically-scattering DM in light of bounds from BBN, CMB, and overproduction. By contrast, fermionic DM absorption on electrons has the potential to be discovered in the current and near-future experiments. 

Fermionic DM absorption by electrons will probe DM masses near their lower bound of a few keV; indeed, if DM is so light, there may be no other way to find it.

\acknowledgments
We thank Benjamin Lehmann, Maxim Pospelov, and Stefano Profumo for helpful discussions. We especially thank Xiao-Dong Ma and Shao-Feng Ge for resolving issues with expressions for the decay rates in a previous version. JD is supported in part by the DOE under contract DE-AC02-05CH11231 and in part by the NSF CAREER grant PHY-1915852. GE is supported by the U.S. Department of Energy, under grant number DE-SC0011637. RM was supported in part by NSF grant PHY-1915314 and the U.S. DoE Contract DE-AC02-05CH11231 and is currently supported by the U.S. DoE under grant de-sc0007859. TTY is supported in part by NSF CAREER grant PHY-1944826. GE thanks the University of Oregon, the Berkeley Center for Theoretical Physics and Lawrence Berkeley National Laboratory for their hospitality during the completion of this work.

\newpage
\onecolumngrid
\appendix
\section{Differential Ionization Cross Section}
\label{app:xsecs}
In this appendix, we build upon the derivation in App.~A of \cite{Essig:2015cda}. The key difference between fermionic absorption cross sections and those of elastic scattering is due to the velocity-independence of the recoil energy (see \cite{Dror:2019dib,Dror:2019onn} for the nuclear target case). In elastic scattering, the incoming DM velocity integral in the averaged differential cross section utilizes the energy-conserving $\delta$ function to impose a physical minimum incoming DM velocity, $v_{\text{min}}$, to achieve a particular $E_R$. 

For instance, in scattering off bound electrons, the relevant piece of the differential scattering rate is (see Eq.(A 12-13) from \cite{Essig:2015cda})
\al{
\int d^3 v g_\chi (\vec{v}) \delta \prn{\Delta E_{1\to 2} + \frac{q^2}{2m_\chi}-qv c \theta_{qv}} 
= \int \frac{dv v^2 d\phi_v}{qv} g_\chi (\vec{v})  \theta \prn{v-v_{\text{min}}} 
= \frac{1}{2q} \eta \prn{v_{\text{min}}}.
}
Note here that $q$ is not evaluated as in the fermionic absorption case, given by Eq.~\eqref{eq:qfrmEcons}. In the second line, we used $\delta \prn{\Delta E_{1\to 2} + \frac{q^2}{2m_\chi}-qv c \theta_{qv}} =\frac{1}{qv} \delta \prn{c \theta_{qv} - c \theta_{qv}^0}$ and noted that $c \theta_{qv}^0=\frac{\Delta E_{1\to 2}}{qv} + \frac{q}{2 m_\chi v}$ which implies $v_{\text{min}}=\frac{\Delta E_{1\to 2}}{q}+\frac{q}{2m_\chi}$. The second line explains the integral form in the first line of Eq.~(A 14) from \cite{Essig:2015cda}. In the last line, the $1/2$ comes from the $c \theta_{qv}$ integral.

By contrast, for fermionic absorption by electrons, the integral over DM's initial velocity is trivial since there is no velocity dependence in the energies at leading order:
\al{
\int d^3 v g_\chi (\vec{v}) \delta \prn{m_\chi +E_B^{nl}-E_R-q} = 1 \cdot \delta \prn{m_\chi +E_B^{nl}-E_R-q}.
}
Thus, the mapping we expect from the usual scattering case is
\al{
\eta \prn{v_{\text{min}}} \longrightarrow 2 q \delta \prn{m_\chi +E_B^{nl}-E_R-q}.
}

With this map, we can write down the differential ionization rate in the case of fermionic absorption. We also need to map the overall factor $\frac{1}{m_\chi^2 m_e^2} \longrightarrow \frac{1}{m_\chi m_e^2 q}$ since the Lorentz-invariant phase space is different for fermionic absorption: the final-states phase space has a $\nu$ instead of an outgoing $\chi$. Using these simple maps to translate the usual DM scattering cross section and starting with Eq.~(A 21) from \cite{Essig:2015cda}, we find
\al{
\frac{d\avg{\sigma^{nl}_{\ion} v}}{dE_R} &= \frac{\Msq}{8 \cdot 16 \pi m_\chi m_e^2} \int \frac{2 q^2}{q} dq \magn{f_\ion \prn{k',q}}^2  \delta \prn{m_\chi +E_B^{nl}-E_R-q} 
&=\frac{\Msq}{64\pi m_\chi m_e^2} \frac{q}{E_R} \magn{f_\ion \prn{k',q}}^2 \theta (q),
}
where $q=m_\chi + E_B^{nl}-E_R$ and $k'=\sqrt{2m_e E_R}$. This is the result quoted in Eq.~\eqref{eq:sigmaion}.

\twocolumngrid
\bibliography{refs.bib}

\end{document}